\documentclass[10pt,conference]{IEEEtran}

\usepackage{amsfonts}
\usepackage{amsmath}
\usepackage{amssymb}
\usepackage{lscape}
\usepackage{epsf}
\usepackage{graphicx}

\newtheorem{theorem}{\indent Theorem}[section]

\newtheorem{EXAMPLE}{\indent Example}[section]

\newcommand{\code}{{\mathcal{C}}}

\newcommand{\cA}{{\mathcal{A}}}
\newcommand{\cD}{{\mathcal{D}}}

\newcommand{\cH}{{\mathcal{H}}}

\newcommand{\cI}{{\mathcal{I}}}
\newcommand{\cJ}{{\mathcal{J}}}

\newcommand{\cQ}{{\mathcal{Q}}}

\newcommand{\bldb}{{\mbox{\boldmath $b$}}}
\newcommand{\bldbb}{{\mbox{\scriptsize \boldmath $b$}}}
\newcommand{\bldc}{{\mbox{\boldmath $c$}}}
\newcommand{\bldcc}{{\mbox{\scriptsize \boldmath $c$}}}

\newcommand{\bldf}{{\mbox{\boldmath $f$}}}
\newcommand{\bldh}{{\mbox{\boldmath $h$}}}

\newcommand{\bldr}{{\mbox{\boldmath $r$}}}
\newcommand{\bldrr}{{\mbox{\scriptsize \boldmath $r$}}}

\newcommand{\bldw}{{\mbox{\boldmath $w$}}}
\newcommand{\bldx}{{\mbox{\boldmath $x$}}}

\newcommand{\bldxi}{{\mbox{\boldmath $\xi$}}}
\newcommand{\bldXi}{{\mbox{\boldmath $\Xi$}}}
\newcommand{\bldG}{{\mbox{\boldmath $G$}}}
\newcommand{\bldL}{{\mbox{\boldmath $L$}}}
\newcommand{\bldy}{{\mbox{\boldmath $y$}}}
\newcommand{\bldyy}{{\mbox{\scriptsize \boldmath $y$}}}
\newcommand{\bldz}{{\mbox{\boldmath $z$}}}

\newcommand{\zeros}{{\mbox{\boldmath $0$}}}
\newcommand{\zeross}{{\mbox{\scriptsize \boldmath $0$}}}

\newcommand{\rrr}{\mathfrak{R}}%
\newcommand{\rrrm}{\rrr^{-}}

\newcommand{\bldzero}{{\mbox{\boldmath $0$}}}

\newcommand{\Prob}{{p}}

    \def\squarebox#1{\hbox to #1{\hfill\vbox to #1{\vfill}}}

\begin{document}

\title{Codeword-Independent Performance of Nonbinary Linear Codes Under Linear-Programming and Sum-Product Decoding}

\author{
\authorblockN{Mark F. Flanagan}
\authorblockA{DEIS/CNIT \\
University of Bologna \\
via Venezia 52, 47023 Cesena (FC), Italy  \\
Email: mark.flanagan@ieee.org}
}

\maketitle

\begin{abstract}
A coded modulation system is considered in which nonbinary coded symbols are mapped directly to nonbinary modulation signals. It is proved that if the modulator-channel combination satisfies a particular symmetry condition, the codeword error rate performance is independent of the transmitted codeword. It is shown that this result holds for both linear-programming decoders and sum-product decoders. In particular, this provides a natural modulation mapping for nonbinary codes mapped to PSK constellations for transmission over memoryless channels such as AWGN channels or flat fading channels with AWGN.   
\end{abstract}

\section{Introduction}
Low-density parity check (LDPC) codes \cite{Gallager}, as well as their nonbinary counterparts \cite{Bennatan} have been shown to exhibit excellent error-correcting performance when decoded by the traditional \emph{sum-product} (SP) decoding algorithm. In \cite{Feldman}, Feldman \emph{et al.} introduced the idea of \emph{linear-programming} (LP) decoding of LDPC codes. This was later generalized to nonbinary codes in \cite{Flanagan}.

For classical coded modulation systems, \emph{geometric uniformity} \cite{Forney} was identified as a symmetry condition which, if satisfied, guarantees codeword error rate performance independent of the transmitted codeword, where maximum-likelihood (ML) decoding is assumed. Some recent coded modulation schemes with SP decoding used this symmetry condition for design \cite{Sridhara}. An analagous symmetry condition was defined in \cite{Richardson} for binary codes over $GF(2)$ with SP decoding; this was extended to nonbinary codes over $GF(q)$ by invoking the concept of \emph{coset} LDPC codes \cite{Bennatan}.

In this work it is shown that for the cases of LP and SP decoding of linear codes over rings, there exists a symmetry condition under which the codeword error rate performance is independent of the transmitted codeword (for the case of LP decoding this theorem generalizes \cite[Theorem 6]{Feldman}, and is stated in \cite{Flanagan}). This provides a condition somewhat akin to geometric uniformity for state-of-the-art nonbinary coded modulation systems. 

\section{General Framework}
We consider codes over finite rings (this includes codes over finite fields, but may be more general). 
Denote by $\rrr$ a ring with $q$ elements, by $0$ its additive identity, and let $\rrrm = \rrr \backslash \{ 0 \}$.   
Let $\code = \{ \bldc \in \rrr^n \; : \; \bldc \cH^T = \bldzero \}$ be a linear code defined with respect to the $m \times n$ parity-check matrix $\cH$ over~$\rrr$. Denote the set of column indices and the set of row indices of $\cH$ by  $\cI = \{1, 2, \cdots, n \}$ 
and $\cJ = \{1, 2, \cdots, m \}$, respectively. 
For $j\in\cJ$, let $\cH_j^{(r)}$ denote the $j$-th row of $\cH$, and for $i\in\cI$, let $\cH_i^{(c)}$ denote the $i$-th column. Denote by $\mbox{supp}(\bldc)$ the support of 
a vector $\bldc$. For each $i\in\cI$, let $\cJ_i = \mbox{supp}(\cH_i^{(c)})$ and for each $j\in\cJ$, let $\cI_j = \mbox{supp}(\cH_j^{(r)})$. 
Also let $\cA_{j,i} = \cI_j\backslash \{i\}$ and $\cD_{j,i} = \cJ_i\backslash \{j\}$.

Given any $\bldc \in \rrr^n$, we say that parity check $j\in\cJ$ is \emph{satisfied} by $\bldc$ if and only if
\begin{equation}
\sum_{i\in\cI_j} c_i \cdot \cH_{j,i} = 0 
\label{eq:parity_check_satisfied}
\end{equation}
For $j\in\cJ$, define the single parity check code $\code_j$ by 
\[
\code_j = \{ ( b_i )_{i\in\cI_j} \; : \; \sum_{i\in\cI_j} b_i \cdot \cH_{j,i} = 0 \}
\]
Note that while the symbols of the codewords in $\code$ are indexed by $\cI$, the symbols of the codewords in $\code_j$ are indexed by $\cI_j$. 
We define the projection mapping for parity check $j\in\cJ$ by
\[
\bldx_j(\bldc) = ( c_i )_{i\in\cI_j}
\]
Then, given any $\bldc \in \rrr^n$, we may say that parity check $j\in\cJ$ is satisfied by $\bldc$ if and only if
\begin{equation}
\bldx_j(\bldc) \in \code_j
\label{eq:word_in_SPC_code}
\end{equation}
since~(\ref{eq:parity_check_satisfied}) and~(\ref{eq:word_in_SPC_code}) are equivalent. 
Also, we say that the vector $\bldc$ is a codeword of $\code$, writing $\bldc \in \code$, if and only if all parity checks $j\in\cJ$ are satisfied by $\bldc$.

Assume that the codeword $\bar{\bldc} = (\bar{c}_1, \bar{c}_2, \cdots, \bar{c}_n) 
\in \code$ has been transmitted over a $q$-ary input
memoryless channel,
and a corrupted word $\bldy = (y_1, y_2, \cdots, y_n) \in \Sigma^n$ has been received. Here $\Sigma$ denotes the set of channel output symbols; we assume that this set either has finite cardinality, or is equal to $\mathbb{R}^l$ or $\mathbb{C}^l$ for some integer $l \ge 1$. In practice, this channel may represent the combination of modulator and physical channel.
It is assumed hereafter that all information words are equally probable, and so all codewords are 
transmitted with equal probability. 

Next we set up some definitions and notation. We define the mapping 
\[
\bldxi \; : \; \rrr \mapsto \{ 0, 1 \}^{q-1} \subset \mathbb{R}^{q-1} 
\]
by 
\[
\bldxi (\alpha) = \bldx = ( x^{(\gamma)} )_{\gamma \in \rrrm } 
\]
such that, for each $\gamma \in \rrrm$,
\[
x^{(\gamma)}=\left\{ \begin{array}{cc}
1 & \textrm{ if } \gamma = \alpha \\
0 & \textrm{ otherwise. }\end{array}\right.
\]
We note that the mapping $\bldxi$ is one-to-one, 
and its image is the set of binary vectors of length $q-1$ with Hamming weight 0 or 1. Building on this, we also define
\[
\bldXi \; : \; \rrr^n \mapsto \{ 0, 1 \}^{(q-1)n} \subset \mathbb{R}^{(q-1)n} 
\]
according to
\[
\bldXi(\bldc) = ( \bldxi(c_1) \; | \; \bldxi(c_2) \; | \; \cdots \; | \; \bldxi(c_n) ) 
\]
We note that $\bldXi$ is also one-to-one.

Now, for vectors $\bldf \in \mathbb{R}^{(q-1)n}$, we adopt the notation
\[
\bldf = ( \bldf_1 \; | \; \bldf_2 \; | \; \cdots \; | \; \bldf_n ) 
\]
where
\[
\forall i \in \cI , \; \bldf_i = ( f_i^{(\alpha)} )_{\alpha \in \rrrm} 
\]
In particular, we define $\boldsymbol{\lambda} \in \mathbb{R}^{(q-1)n}$ by setting, for each $i\in\cI$, $\alpha\in\rrrm$,
\[
\lambda_i^{(\alpha)} = \log \left( \frac{\Prob ( y_i | 0 )}{ \Prob ( y_i | \alpha ) } \right) 
\]
and $p(y_i|c_i)$ denotes the channel output probability (density) conditioned on the channel input. 

Also, we may use this notation to write the inverse of $\bldXi$ as
\[
\bldXi^{-1} (\bldf) = ( \bldxi^{-1}(\bldf_1), \bldxi^{-1}(\bldf_2), \cdots, \bldxi^{-1}(\bldf_n) ) 
\]

\section{Decoding Algorithms}
\subsection{Linear-Programming Decoder}
The linear-programming (LP) decoder of \cite{Flanagan} operates as follows. The linear program described here is equivalent to that given in \cite{Flanagan}; however, some changes of notation have been made in order to facilitate the proof to come in section \ref{sec:main_result}. The variables of the LP are 
\[
f_i^{(\alpha)} \; \mbox{ for each } \;  i\in\cI, \alpha \in \rrrm
\]
and 
\[
w_{j,\bldbb} \; \mbox{ for each } \;  j\in\cJ, \bldb \in C_j 
\]
and the constraints are

\begin{eqnarray}
\forall j \in \cJ, \; \forall \bldb \in C_j,  \quad  w_{j,\bldbb} \ge 0 
\label{eq:equation-polytope-3} 
\end{eqnarray} 
and
\begin{equation}
\forall j \in \cJ, \quad \sum_{\bldbb \in C_j} w_{j,\bldbb} = 1 
\label{eq:equation-polytope-4} 
\end{equation} 
and
\begin{eqnarray}
&  \forall j \in \cJ, \; \forall i \in \cI_j, \; \forall \alpha \in \rrrm, \nonumber \\
& f_i^{(\alpha)} = 
\sum_{\bldbb \in C_j, \; b_i=\alpha} w_{j,\bldbb} 
\label{eq:equation-polytope-5} 
\end{eqnarray} 
The set of points $(\bldf,\bldw)$ which satisfy~(\ref{eq:equation-polytope-3})-(\ref{eq:equation-polytope-5}) form a polytope denoted by $\cQ$. The cost function to be minimized over this polytope is $ F(\bldf) = \boldsymbol{\lambda} \bldf^T $, and the minimizer is denoted by $\hat{\bldf}$. If $\hat{\bldf} \in \{0,1\}^{(q-1)n}$, the output is the codeword $\bldXi^{-1}(\hat{\bldf})$ (it is proved in \cite{Flanagan} that this must be the maximum-likelihood codeword). Otherwise, the decoder outputs a `decoding failure'. 
\subsection{Sum-Product Decoder}
The sum-product (SP) decoder operates as follows. Note that in practice, computations are usually carried out in the log-domain, but this does not affect our analysis.

Initializing
\begin{equation}
m_i(\alpha) = p(y_i|\alpha) \;\;\;\;\; \forall i\in\cI, \; \forall \alpha \in \rrr
\label{eq:SPA_init_channel}
\end{equation}
and 
\begin{equation}
m_{j,i}^{D,0}(\alpha) = 1 \;\;\;\;\; \forall j\in\cJ, \; \forall i \in \cI_j, \; \forall \alpha \in \rrr
\label{eq:SPA_init_down}
\end{equation}
$N$ iterations of fully parallel SP decoding may be represented by the following recursive formulas. For each $k=1,2,\cdots N$,
\begin{equation}
m_{j,i}^{U,k}(\alpha) = m_i(\alpha) \cdot \prod_{l\in \cD_{j,i}} m_{l,i}^{D,k-1}(\alpha)
\label{eq:SPA_up}
\end{equation}
for each $j\in\cJ$, $i\in\cI_j$, $\alpha\in\rrr$, and
\begin{equation}
m_{j,i}^{D,k}(\alpha) = \sum_{\sum_{l\in \cA_{j,i}} d_l \cH_{j,l} = -\alpha \cH_{j,i}} \left\{ \prod_{l\in\cA_{j,i}} m_{j,l}^{U,k}(d_l) \right\}
\label{eq:SPA_down}
\end{equation}
for each $j\in\cJ$, $i\in\cI_j$, $\alpha\in\rrr$. Finally, decisions are made via
\begin{equation}
g_i(\alpha) = m_i(\alpha) \cdot \prod_{j\in\cJ_i} m_{j,i}^{D,N}(\alpha) \; \forall i \in \cI, \; \forall \alpha \in \rrr
\label{eq:SPA_summaries}
\end{equation}
and
\begin{equation}
h_i = \arg \max_{\alpha\in\rrr} \left\{ g_i(\alpha) \right\} \; \forall i\in\cI
\label{eq:SPA_decisions}
\end{equation}
The output of the decoder is then $\bldh = (h_1, h_2, \cdots, h_n)$. 

\section{Main Result\label{sec:main_result}}

{\bf Symmetry Condition.}  

For each $\beta\in\rrr$, there exists a bijection 
\[
\tau_{\beta} \; : \; \Sigma \longrightarrow \Sigma 
\]
such that the channel output probability (density) conditioned on the channel input satisfies
\begin{equation}
p(y|\alpha) = p(\tau_{\beta}(y)|\alpha-\beta) 
\label{eq:symmetry_condition}
\end{equation}
for all $y \in \Sigma$, $\alpha \in \rrr$.
When $\Sigma$ is equal to $\mathbb{R}^l$ or $\mathbb{C}^l$ for $l \ge 1$, the mapping $\tau_{\beta}$ is assumed to be isometric with respect to Euclidean distance in $\Sigma$, for every $\beta\in\rrr$.

In the following, \emph{codeword error} is defined as the event where the decoder output is not equal to the transmitted codeword.

\begin{theorem}
Under the stated symmetry condition, the probability of codeword error is independent of the transmitted codeword

(a) under linear-programming decoding

(b) under sum-product decoding.

\label{htrm:equl-prob} 
\end{theorem}

\begin{proof}
We shall prove the theorem for the case where $\Sigma$ has infinite cardinality; the case of discrete $\Sigma$ may be handled similarly. Fix some codeword $\bldc \in \code$, $\bldc \ne \zeros$. We wish to prove that
\[
\mbox{Pr}(\mbox{Err} \; | \; \bldc) = \mbox{Pr}(\mbox{Err} \; | \; \zeros)
\]
where $\mbox{Pr}(\mbox{Err} \; | \; \bldc)$ denotes the probability of codeword error given that the codeword $\bldc$ was transmitted. 

Now
\[
\mbox{Pr}(\mbox{Err} \; | \; \bldc) = \mbox{Pr}(\bldy \in B(\bldc) \; | \; \bldc)
\]
where $B(\bldc)$ is the set of all receive words which may cause codeword error, given that $\bldc$ was transmitted. Also
\[
\mbox{Pr}(\mbox{Err} \; | \; \zeros) = \mbox{Pr}(\bldy \in B(\zeros) \; | \; \zeros)
\]
So we write
\begin{equation}
\mbox{Pr}(\mbox{Err} \; | \; \bldc) = \int_{\bldyy \in B(\bldcc)} \Prob ( \; \bldy \; | \; \bldc \; ) \; d\bldy
\label{eq:integral_1}
\end{equation}
and
\begin{equation}
\mbox{Pr}(\mbox{Err} \; | \; \zeros) = \int_{\tilde{\bldyy} \in B(\zeross)} \Prob ( \; \tilde{\bldy} \; | \; \zeros \; ) \; d\tilde{\bldy}
\label{eq:integral_2}
\end{equation}
Now, setting $\alpha=\beta$ in the symmetry condition~(\ref{eq:symmetry_condition}) yields
\begin{equation}
p(y|\beta) = p(\tau_{\beta}(y)|0)
\label{eq:equality_in_symmetry_condition}
\end{equation}
for any $y \in \Sigma$, $\beta \in \rrr$.

We now define $\tilde{\bldy} = \bldG(\bldy)$ as follows. For every $i\in\cI$, if $c_i=\beta \in \rrr$ then 
\[
\tilde{y}_i = \tau_{\beta}(y_i)
\]
We note that $\bldG$ is a bijection from the set $\Sigma^n$ to itself, and that if $\bldy, \bldz \in \Sigma^n$ and $c_i=\beta\in\rrr$ then 
\[
\| y_i - z_i \| ^2 = \| \tau_{\beta}(y_i) - \tau_{\beta}(z_i) \| ^2
\] 
and so
\[
\| \bldG(\bldy) - \bldG(\bldz) \| ^2 = \| \bldy - \bldz \| ^2
\] 
i.e. $\bldG$ is isometric with respect to Euclidean distance in $\Sigma^n$.

We prove that the integral~(\ref{eq:integral_1}) may be transformed to~(\ref{eq:integral_2}) via the substitution $\tilde{\bldy} = \bldG(\bldy)$. First, we have
\begin{eqnarray*}
\Prob ( \; \bldy \; | \; \bldc \; ) & = & \prod_{i\in\cI} \Prob ( y_i | c_i ) \\
 & = & \prod_{\beta\in\rrr} \prod_{i\in\cI , c_i=\beta} \Prob ( y_i | \beta ) \\
 & = & \prod_{\beta\in\rrr} \prod_{i\in\cI, c_i=\beta} \Prob ( \tau_{\beta}(y_i) | 0 ) \\
 & = & \prod_{\beta\in\rrr} \prod_{i\in\cI, c_i=\beta} \Prob ( \tilde{y}_i | 0 ) \\
 & = & \prod_{i\in\cI} \Prob ( \tilde{y}_i | 0 ) \\
 & = & \Prob ( \; \tilde{\bldy} \; | \; \zeros \; )
\end{eqnarray*}
Since $\bldG$ is isometric with respect to Euclidean distance in $\Sigma^n$, it follows that the Jacobian determinant of the transformation is equal to unity. Therefore, to complete the proof, we need only show that 
\[
\bldy \in B(\bldc) \mbox{ if and only if } \tilde{\bldy} \in B(\zeros)
\]
We prove this separately for the two cases of linear-programming and sum-product decoding.

(a) \emph{Under linear-programming decoding:}

Here 
\begin{equation*}
\begin{split}
B(\bldc) = \{ \bldy \in \Sigma^n \; : \; \exists (\bldf, \bldw) & \in \cQ, \bldf \ne \bldXi(\bldc) \\
 & \mbox{with } \boldsymbol{\lambda} \bldf ^T \le \boldsymbol{\lambda} \bldXi(\bldc) ^T \}
\end{split}
\end{equation*} 
Recall that here $\boldsymbol{\lambda}$ is a function of $\bldy$ via
\begin{equation}
\lambda_i^{(\alpha)} = \log \left( \frac{\Prob ( y_i | 0 )}{ \Prob ( y_i | \alpha ) } \right) 
\label{eq:lambda_def}
\end{equation}
for $i\in\cI$, $\alpha\in\rrrm$. Also
\begin{equation*}
\begin{split}
B(\zeros) = \{ \tilde{\bldy} \in \Sigma^n \; : \; \exists (\tilde{\bldf}, \tilde{\bldw}) & \in \cQ, \tilde{\bldf} \ne \bldXi(\zeros) \\
 & \mbox{with } \tilde{\boldsymbol{\lambda}} \tilde{\bldf} ^T \le \tilde{\boldsymbol{\lambda}} \bldXi(\zeros) ^T \} 
\end{split}
\end{equation*}
Here $\tilde{\boldsymbol{\lambda}}$ is a function of $\tilde{\bldy}$ via
\begin{equation}
\tilde{\lambda}_i^{(\alpha)} = \log \left( \frac{\Prob ( \tilde{y}_i | 0 )}{ \Prob ( \tilde{y}_i | \alpha ) } \right) 
\label{eq:lambda_tilde_def}
\end{equation}
for $i\in\cI$, $\alpha\in\rrrm$. 
We begin by relating the elements of $\boldsymbol{\lambda}$ (defined by~(\ref{eq:lambda_def})) to the elements of $\tilde{\boldsymbol{\lambda}}$ (defined by~(\ref{eq:lambda_tilde_def})). Let $i\in\cI$, $\alpha\in\rrrm$. Suppose $c_i=\beta \in \rrr$. We then have
\begin{eqnarray*}
\lambda_i^{(\alpha)} & = & \log \left( \frac{\Prob ( y_i | 0 )}{ \Prob ( y_i | \alpha ) } \right) \\
 & = & \log \left( \frac{\Prob ( \tau_{\beta}(y_i) | -\beta )}{ \Prob ( \tau_{\beta}(y_i) | \alpha-\beta ) } \right) \\
 & = & \log \left( \frac{\Prob ( \tilde{y}_i | -\beta )}{ \Prob ( \tilde{y}_i | \alpha-\beta ) } \right) 
\end{eqnarray*}
This yields
\[
\lambda_i^{(\alpha)} =\left\{ \begin{array}{ccc}
\tilde{\lambda}_i^{(\alpha)} & \textrm{ if } \beta=0 \\
-\tilde{\lambda}_i^{(-\alpha)} & \textrm{ if } \alpha=\beta \\
\tilde{\lambda}_i^{(\alpha-\beta)}-\tilde{\lambda}_i^{(-\beta)} & \textrm{ otherwise. }\end{array}\right.
\]

Next, for any point $(\bldf,\bldw)\in\cQ$ we define a new point $(\tilde{\bldf},\tilde{\bldw})$ as follows. For all $i\in\cI$, $\alpha \in \rrrm$, if $c_i=\beta \in \rrr$ then
\begin{equation}
\tilde{f}_i^{(\alpha)} = \left\{ \begin{array}{ccc}
1 - \sum_{\gamma\in\rrrm} f_i^{(\gamma)} & \textrm{ if } \alpha=-\beta \\
f_i^{(\alpha+\beta)} & \textrm{ otherwise. }\end{array}\right. 
\label{eq:f_to_ftilde}
\end{equation}
For all $j\in\cJ$, $\bldr \in \code_j$ we define
\[
\tilde{w}_{j,\bldrr} = w_{j,\bldbb}
\]
where
\[
\bldb = \bldr + \bldx_j(\bldc)
\]

Next we prove that for every $(\bldf,\bldw) \in \cQ$, the new point $(\tilde{\bldf},\tilde{\bldw})$ lies in $\cQ$ and thus is a feasible solution for the LP. Constraints~(\ref{eq:equation-polytope-3}) and~(\ref{eq:equation-polytope-4}) obviously hold from the definition of $\tilde{\bldw}$. To verify~(\ref{eq:equation-polytope-5}), we let $j\in\cJ$, $i\in\cI_j$ and $\alpha \in \rrrm$. We also let $c_i=\beta\in \rrr$. We now check two cases: 
\begin{itemize}
\item
If $\alpha=-\beta$,
\begin{eqnarray*}
\tilde{f}_i^{(\alpha)} & = & 1 - \sum_{\gamma\in\rrrm} f_i^{(\gamma)} \\
 & = & \sum_{\bldbb \in \code_j} w_{j,\bldbb} - \sum_{\gamma\in\rrrm} \sum_{\bldbb \in \code_j, \; b_i=\gamma} w_{j,\bldbb} \\
 & = & \sum_{\bldbb \in \code_j, \; b_i=0} w_{j,\bldbb} \\
 & = & \sum_{\bldrr \in \code_j, \; r_i=\alpha} \tilde{w}_{j,\bldrr}
\end{eqnarray*}
\item
If $\alpha \ne -\beta$,
\begin{eqnarray*}
\tilde{f}_i^{(\alpha)} = f_i^{(\alpha+\beta)} & = & \sum_{\bldbb \in \code_j, \; b_i=\alpha+\beta} w_{j,\bldbb} \\
 & = & \sum_{\bldrr \in \code_j, \; r_i=\alpha} \tilde{w}_{j,\bldrr}
\end{eqnarray*}
\end{itemize}
Therefore $(\tilde{\bldf},\tilde{\bldw}) \in \cQ$, i.e. $(\tilde{\bldf},\tilde{\bldw})$ is a feasible solution for the LP. We write $(\tilde{\bldf},\tilde{\bldw}) = \bldL(\bldf,\bldw)$. We also note that the mapping $\bldL$ is a bijection from $\cQ$ to itself; this is easily shown by verifying the inverse
\begin{equation}
f_i^{(\alpha)} = \left\{ \begin{array}{ccc}
1 - \sum_{\gamma\in\rrrm} \tilde{f}_i^{(\gamma)} & \textrm{ if } \alpha=\beta \\
\tilde{f}_i^{(\alpha-\beta)} & \textrm{ otherwise }\end{array}\right. 
\label{eq:ftilde_to_f}
\end{equation}
for all $i\in\cI$, $\alpha \in \rrrm$, and 
\[
w_{j,\bldbb} = \tilde{w}_{j,\bldrr} 
\]
where
\[
\bldr = \bldb - \bldx_j(\bldc)
\]
for all $j\in\cJ$, $\bldb \in \code_j$.

We now prove that for every $(\bldf,\bldw) \in \cQ$, $(\tilde{\bldf},\tilde{\bldw}) = \bldL(\bldf,\bldw)$ satisfies
\begin{equation}
\boldsymbol{\lambda} \bldf ^T - \boldsymbol{\lambda} \bldXi(\bldc) ^T = \tilde{\boldsymbol{\lambda}} \tilde{\bldf} ^T - \tilde{\boldsymbol{\lambda}} \bldXi(\zeros) ^T  
\label{eq:relative_cost_fn_1}
\end{equation}
We achieve this by proving
\begin{equation}
\boldsymbol{\lambda}_i \bldf_i ^T - \boldsymbol{\lambda}_i \bldxi(c_i) ^T = \tilde{\boldsymbol{\lambda}}_i \tilde{\bldf}_i ^T - \tilde{\boldsymbol{\lambda}}_i \bldxi(0) ^T  
\label{eq:relative_cost_fn_2}
\end{equation}
for every $i\in \cI$. We may then obtain~(\ref{eq:relative_cost_fn_1}) by summing~(\ref{eq:relative_cost_fn_2}) over $i\in\cI$. Let $c_i=\beta \in \rrr$. We consider two cases:
\begin{itemize}
\item
If $\beta=0$,~(\ref{eq:relative_cost_fn_2}) becomes
\[
\boldsymbol{\lambda}_i \bldf_i ^T = \tilde{\boldsymbol{\lambda}}_i \tilde{\bldf}_i ^T 
\]
which holds since in this case $\tilde{\lambda}_i^{(\alpha)} = \lambda_i^{(\alpha)}$ and $\tilde{f}_i^{(\alpha)} = f_i^{(\alpha)}$ for all $\alpha\in\rrrm$.
\item
If $\beta \ne 0$,
\begin{equation*}
\begin{split}
& \boldsymbol{\lambda}_i \bldf_i ^T - \boldsymbol{\lambda}_i \bldxi(c_i) ^T = \sum_{\gamma\in\rrrm} \lambda_i^{(\gamma)} f_i^{(\gamma)} - \lambda_i^{(\beta)} \\
 & = \sum_{\substack{\gamma\in\rrrm \\ \gamma \ne \beta}} \left( \tilde{\lambda}_i^{(\gamma-\beta)} - \tilde{\lambda}_i^{(-\beta)} \right)  f_i^{(\gamma)} - \tilde{\lambda}_i^{(-\beta)} f_i^{(\beta)} + \tilde{\lambda}_i^{(-\beta)} \\
 & = \sum_{\substack{\alpha\in\rrrm \\ \alpha \ne -\beta}} \tilde{\lambda}_i^{(\alpha)} f_i^{(\alpha+\beta)} + \tilde{\lambda}_i^{(-\beta)} \left( 1 - \sum_{\gamma\in\rrrm} f_i^{(\gamma)} \right) \\
 & = \sum_{\alpha\in\rrrm} \tilde{\lambda}_i^{(\alpha)} \tilde{f}_i^{(\alpha)} \\
 & = \tilde{\boldsymbol{\lambda}}_i \tilde{\bldf}_i ^T - \tilde{\boldsymbol{\lambda}}_i \bldxi(0) ^T
\end{split}
\end{equation*}
\end{itemize}
where we have made use of the substitution $\alpha = \gamma - \beta$ in the third line. Therefore~(\ref{eq:relative_cost_fn_2}) holds, proving~(\ref{eq:relative_cost_fn_1}).

Finally, we note that it is easy to show, using~(\ref{eq:f_to_ftilde}) and~(\ref{eq:ftilde_to_f}), that $\bldf = \bldXi(\bldc)$ if and only if $\tilde{\bldf} = \bldXi(\zeros)$.

Putting together these results, we may make the following statement. Suppose we are given $\bldy, \tilde{\bldy} \in \Sigma^n$ with $\tilde{\bldy} = \bldG(\bldy)$. Then the point $(\bldf,\bldw) \in \cQ$ satisfies $\bldf \ne \bldXi(\bldc)$ and $\boldsymbol{\lambda} \bldf ^T \le \boldsymbol{\lambda} \bldXi(\bldc) ^T$ if and only if the point $(\tilde{\bldf},\tilde{\bldw}) = \bldL(\bldf,\bldw) \in \cQ$ satisfies $\tilde{\bldf} \ne \bldXi(\zeros)$ and $\tilde{\boldsymbol{\lambda}} \tilde{\bldf} ^T \le \tilde{\boldsymbol{\lambda}} \bldXi(\zeros) ^T$. This statement, along with the fact that both $\bldG$ and $\bldL$ are bijective, proves that 
\[
\bldy \in B(\bldc) \mbox{ if and only if } \tilde{\bldy} \in B(\zeros)
\]
This completes the proof of the theorem for the case of LP decoding.

(b) \emph{Under Sum-Product Decoding:}

Recall that all decoder variables appearing in equations (\ref{eq:SPA_init_channel})-(\ref{eq:SPA_decisions}) are functions of $\bldy$ via (\ref{eq:SPA_init_channel}). For any such variable $x$, let $\tilde{x}$ denote the corresponding variable with $\tilde{\bldy}$ as input. Then we have, for all $i\in\cI$, $\alpha\in\rrr$, where $c_i=\beta$,
\begin{eqnarray*}
m_i(\alpha) = p(y_i|\alpha) = p(\tau_{\beta}(y_i)|\alpha-\beta) \\
= p(\tilde{y}_i|\alpha-c_i) = \tilde{m}_i(\alpha-c_i) 
\end{eqnarray*}
Next we prove by induction that for all $k=0,1,\cdots N$,
\begin{equation}
m_{j,i}^{D,k}(\alpha) = \tilde{m}_{j,i}^{D,k}(\alpha-c_i)
\label{formula_for_induction}
\end{equation}
for all $j\in\cJ$, $i\in\cI_j$, $\alpha\in\rrr$. This result holds for the base case $k=0$ because from (\ref{eq:SPA_init_down})
\[
m_{j,i}^{D,0}(\alpha) = \tilde{m}_{j,i}^{D,0}(\alpha) = 1 \;\;\;\;\; \forall j\in\cJ, \; \forall i \in \cI_j, \; \forall \alpha \in \rrr
\]
Assuming that (\ref{formula_for_induction}) holds for some $k=r-1\in\{ 0,1,\cdots N-1\}$ (and for all $j\in\cJ$, $i\in\cI_j$, $\alpha\in\rrr$), we obtain by (\ref{eq:SPA_up})  
\begin{eqnarray*}
m_{j,i}^{U,r}(\alpha) & = & m_i(\alpha) \cdot \prod_{l\in \cD_{j,i}} m_{l,i}^{D,r-1}(\alpha) \\
& = & \tilde{m}_i(\alpha-c_i) \cdot \prod_{l\in \cD_{j,i}} \tilde{m}_{l,i}^{D,r-1}(\alpha-c_i) \\
& = & \tilde{m}_{j,i}^{U,r}(\alpha-c_i) 
\end{eqnarray*}
for all $j\in\cJ$, $i\in\cI_j$, $\alpha\in\rrr$. So, by (\ref{eq:SPA_down}),
\begin{eqnarray*}
m_{j,i}^{D,r}(\alpha) & = & \sum_{\sum_{l\in \cA_{j,i}} d_l \cH_{j,l} = -\alpha \cH_{j,i}} \left\{ \prod_{l\in\cA_{j,i}} m_{j,l}^{U,r}(d_l) \right\} \\
& = & \sum_{\sum_{l\in \cA_{j,i}} d_l \cH_{j,l} = -\alpha \cH_{j,i}} \left\{ \prod_{l\in\cA_{j,i}} \tilde{m}_{j,l}^{U,r}(d_l-c_l) \right\} \\
& = & \sum_{\sum_{l\in \cA_{j,i}} b_l \cH_{j,l} = -(\alpha-c_i) \cH_{j,i}} \left\{ \prod_{l\in\cA_{j,i}} \tilde{m}_{j,l}^{U,r}(b_l) \right\} \\
& = & \tilde{m}_{j,i}^{D,r}(\alpha-c_i) 
\end{eqnarray*}
for all $j\in\cJ$, $i\in\cI_j$, $\alpha\in\rrr$, where we have made the substitution $b_l=d_l-c_l$ for each $l\in\cI_j$, and used the fact that $\sum_{l\in \cA_{j,i}} c_l \cH_{j,l} = -c_i \cH_{j,i}$ since $c\in\code$.
It follows by the principle of induction that (\ref{formula_for_induction}) holds for every $k=0,1,\cdots N$, $j\in\cJ$, $i\in\cI_j$, $\alpha\in\rrr$. Therefore by (\ref{eq:SPA_summaries})
\begin{eqnarray*}
g_i(\alpha) & = & m_i(\alpha) \cdot \prod_{j\in\cJ_i} m_{j,i}^{D,N}(\alpha) \\ 
& = & \tilde{m}_i(\alpha-c_i) \cdot \prod_{j\in\cJ_i} \tilde{m}_{j,i}^{D,N}(\alpha-c_i) \\
& = & \tilde{g}_i(\alpha-c_i) 
\end{eqnarray*}
for all $i\in\cI$, $\alpha\in\rrr$, and so by (\ref{eq:SPA_decisions}), $\tilde{h}_i = h_i-c_i$ for all $i\in \cI$. Therefore $\bldh \ne \bldc$ if and only if $\tilde{\bldh} \ne \zeros$. We conclude that 
\[
\bldy \in B(\bldc) \mbox{ if and only if } \tilde{\bldy} \in B(\zeros)
\]
This completes the proof of the theorem for the case of SP decoding. It is trivial to see that this proof generalizes to the case of optional early exit of the iterative loop on successful completion of a syndrome check.
\end{proof}

\section{Application: Nonbinary Codes Mapped to PSK modulation}

While this theorem may be shown to apply to other coded modulation systems such as nonbinary coded orthogonal modulation over memoryless channels and nonbinary coding over the discrete memoryless $q$-ary symmetric channel, we focus in this paper on the practical application of nonbinary codes mapped directly to PSK symbols and transmitted over a memoryless channel. Here $\Sigma = \mathbb{C}$, and denoting the ring elements by $\rrr = \{ a_0, a_1, \cdots, a_{q-1} \}$, the modulation mapping may be written without loss of generality as
\[
M \; : \; \rrr \mapsto \mathbb{C} 
\]
such that
\begin{equation}
M (a_k) = \exp \left( \frac{\imath 2 \pi k}{q} \right) 
\label{eq:modulation_mapping}
\end{equation}
for $k = 0, 1, \cdots, q-1$ (here $\imath=\sqrt{-1}$). Here~(\ref{eq:equality_in_symmetry_condition}), together with the rotational symmetry of the $q$-ary PSK constellation, motivates us to define, for every $\beta= a_k \in\rrr$,
\begin{equation}
\tau_\beta(x) = \exp \left( \frac{-\imath 2 \pi k}{q} \right) \cdot x \qquad \forall x \in \mathbb{C} 
\label{eq:tau_defn_PSK}
\end{equation}
Next, we also impose the condition that $\rrr$ under addition is a cyclic group. To see why we impose this condition, let $\alpha = a_k \in\rrr$ and $\beta = a_l \in\rrr$. By the symmetry condition we must have
\[
p(y_i|\alpha+\beta) = p(\tau_{\alpha+\beta}(y_i)|0) 
\]
and also
\begin{equation*}
p(y_i|\alpha+\beta) = p(\tau_{\beta}(y_i)|\alpha) = p(\tau_{\alpha}(\tau_{\beta}(y_i))|0) 
\end{equation*}

In order to equate these two expressions, we impose the condition $\tau_{\alpha+\beta}(x) = \tau_{\alpha}(\tau_{\beta}(x))$ for all $x \in \mathbb{C}$, $\alpha,\beta \in\rrr$. Letting $\alpha+\beta = a_p\in\rrr$, and using~(\ref{eq:tau_defn_PSK}) yields
\[
\exp \left( \frac{- \imath 2 \pi k}{q} \right) \cdot \exp \left( \frac{- \imath 2 \pi l}{q} \right) = \exp \left( \frac{- \imath 2 \pi p}{q} \right)
\] 
and thus $p\equiv k+l \mod q$. 

Therefore, we must have 
\begin{equation}
a_k + a_l = a_{(k+l \!\!\!\! \mod q)}
\label{eq:cyclic_group}
\end{equation} 
for all $a_k, a_l \in \rrr$. This implies that $\rrr$, under addition, is a cyclic group.

It is easy to check that the condition that $\rrr$ under addition is cyclic, encapsulated by~(\ref{eq:cyclic_group}), along with the modulation mapping (\ref{eq:modulation_mapping}), satisfies the symmetry condition, where the appropriate mappings $\tau_{\beta}$ are given by~(\ref{eq:tau_defn_PSK}). This means that codeword-independent performance is guaranteed for such systems using nonbinary codes with PSK modulation. This applies to AWGN, flat fading wireless channels, and OFDM systems transmitting over frequency selective channels with sufficiently long cyclic prefix.
\section*{Acknowledgment}
The author would like to thank M. Greferath and V. Skachek for providing helpful comments which improved the presentation of this paper. This work was supported by the Claude Shannon Institute, Dublin, Ireland (Science Foundation Ireland Grant 06/MI/006), the University of Bologna (ESRF-ISA) and the EC-IST Optimix project (IST-214625).


%

\end{document}